\documentclass[
]{ceurart}

\usepackage{listings}
\usepackage{acro}
\usepackage{cleveref}
\usepackage{graphicx}
\usepackage{hyperref}
\usepackage[export]{adjustbox}
\usepackage[normalem]{ulem}
\useunder{\uline}{\ul}{}
\DeclareAcronym{ner}{
  short = NER,
  long  = named entity recognition
}
\DeclareAcronym{nel}{
  short = NEL,
  long  = named entity linking
}
\DeclareAcronym{dh}{
  short = DH,
  long  = Digital Humanities
}

\newcommand{\eg}{\emph{e.g., }}
\newcommand{\ie}{{i.e., }}

\begin{document}

\copyrightyear{2025}
\copyrightclause{Copyright for this paper by its authors.
  Use permitted under Creative Commons License Attribution 4.0
  International (CC BY 4.0).}

\conference{Human-Computer Interaction Slovenia 2025, October 13, 2025, Koper, Slovenia}

\title{NERVIS: An Interactive System for Graph-Based Exploration and Editing of Named Entities}



\author[]{Uroš Šmajdek}[%
email=uros.smajdek@fri.uni-lj.si,
]
\fnmark[1]

\author[]{Ciril Bohak}[%
orcid=0000-0002-9015-2897,
email=ciril.bohak@fri.uni-lj.si,
url=https://lgm.fri.uni-lj.si/ciril,
]
\fnmark[1]
\cormark[1]

\address[]{University of Ljubljana, Faculty of Computer and Information Science, Večna pot 113, SI-1000 Ljubljana, Slovenia}

\cortext[1]{Corresponding author.}
\fntext[1]{These authors contributed equally.}

\begin{abstract}
We present an interactive visualization system for exploring named entities and their relationships across document collections. The system is designed around a graph-based representation that integrates three types of nodes: documents, entity mentions, and entities. Connections capture two key relationship types: (i) identical entities across contexts, and (ii) co-locations of mentions within documents. Multiple coordinated views enable users to examine entity occurrences, discover clusters of related mentions, and explore higher-level entity group relationships. To support flexible and iterative exploration, the interface offers fuzzy views with approximate connections, as well as tools for interactively editing the graph by adding or removing links, entities, and mentions, as well as editing entity terms. Additional interaction features include filtering, mini-map navigation, and export options to JSON or image formats for downstream analysis and reporting. This approach contributes to human-centered exploration of entity-rich text data by combining graph visualization, interactive refinement, and adaptable perspectives on relationships. The deployment of the system is available at: \url{https://ul-fri-lgm.github.io/NERVIS/}.
\end{abstract}

\begin{keywords}
  Named Entity Visualization \sep
  Graph Exploration \sep
  Interactive Visualization
\end{keywords}

\maketitle

\section{Introduction}
\label{sec:intro}
Textual data continues to grow at an unprecedented scale across domains such as scientific publishing, journalism, and social media, but also with the digitization of historical text corpora. Extracting meaning from these large corpora requires more than keyword search or frequency analysis; it requires understanding the entities mentioned in the text and the relationships between them. \ac{ner} and related information extraction methods provide a way to identify people, organizations, locations, and other key entities. However, the resulting collections of mentions and links are often complex, redundant, and difficult to navigate without appropriate visualization and interaction techniques.

Graph-based visualization offers a natural way to represent entities and their relationships. Nodes can represent documents, entity mentions, or consolidated entities, while edges capture relationships such as co-location of mentions or equivalence of entities across contexts. Yet, static graph layouts alone are insufficient for the exploration of large and ambiguous entity networks. What is needed are interactive approaches that allow users to flexibly explore different levels of granularity, filter and refine connections, and iteratively adjust the structure of the graph to reflect domain knowledge and emerging insights.

In this paper, we introduce an interactive visualization system for entity-centric text exploration. The system combines graph representations with multiple coordinated views, supporting fuzzy and precise connections, user-driven graph editing, and export capabilities for integration with downstream analysis pipelines. By emphasizing exploratory flexibility and human-centered control, our approach bridges automatic information extraction with interactive sensemaking, enabling users to uncover patterns and relationships in complex entity-rich text collections.
\section{Related Work}
\label{sec:related}

A number of prior works provide foundations for designing interactive visualization systems that support exploration of large and complex text collections. Paul et al.~\cite{Paul2018} introduce TexTonic, a visual analytics system that enables exploration through hierarchical clustering and direct manipulation of terms. Their approach highlights how interactive interfaces can support users in dynamically engaging with data to discover patterns and relationships among entities and mentions. In a complementary direction, Lee et al.~\cite{Lee2012} expand design considerations for information visualization interactions beyond the traditional mouse and keyboard. Their work underscores the importance of multimodal interaction, which can broaden the accessibility and flexibility of visualization systems.

Scalability and performance are equally critical aspects of interactive visualization. Tao et al.~\cite{Tao2019} present Kyrix, a system designed for large-scale web-based visualization, demonstrating techniques for efficient pan and zoom interactions at scale. Their emphasis on performance optimization ensures responsiveness when managing very large datasets, an aspect directly relevant to maintaining smooth exploration of entity-rich corpora. Such techniques align well with our use of fuzzy views and approximate connections to facilitate exploratory analysis without sacrificing interactivity.

The need for robust navigation and filtering has also been well established. Al Nasar et al.~\cite{AlNasar2011} discuss principles of interactive personal information management systems, emphasizing personalized interfaces to help users manage large information collections. While their focus is on photographs and videos, the principles are transferable to text-based entity exploration. Similarly, Sedig and Parsons~\cite{Sedig2013} propose a pattern-based framework for interaction design that supports complex cognitive activities, providing conceptual guidance for structuring interactions within entity exploration systems.

Editing and refinement capabilities form another important dimension of related work. Satyanarayan et al.~\cite{Satyanarayan2016} propose Reactive Vega, a declarative framework for specifying interactive visualizations. Their approach demonstrates how flexible editing, specification, and interaction can empower users to tailor visualizations to their analytical needs. This perspective informs our focus on enabling users to add, remove, and refine entities and connections within the graph representation.

Recent work has advanced domain-specific analytical systems enhanced with graph-based visualizations. Rahman et al.~\cite{Rahman2021} present DiaVis, a dashboard for iterative exploration of diabetes-related data, highlighting the benefits of refinement cycles for sustaining engagement—an insight we adopt for entity-level analysis. Heberle et al.~\cite{Heberle2017} propose a web-based system for biological networks that employs rule-based filtering and automated layout generation to support scalable exploration. Similarly, Kwak et al.~\cite{Kwak2023} develop a visual analytics framework for stroke care networks, using multiple coordinated views to capture patterns across different levels of analysis, from national to community scale. In cybersecurity, a recent system by Razbelj et al.~\cite{Rabzelj2023} introduces a graph-based approach to visualizing honeypot-captured attacks, enabling interactive exploration of large, heterogeneous datasets to detect recurring patterns and uncover hidden connections. Collectively, these systems demonstrate how interactive visualization, filtering, and scalable layouts can support domain-specific analysis.

Finally, Mongiov\`{i} and Gangemi~\cite{Mongiovi2024} propose GRAAL, a graph-based retrieval system for collecting related passages across multiple documents. By emphasizing semantic interactions and subgraph representations, GRAAL directly addresses the challenge of identifying co-occurring entity mentions in large text collections. Their methodology provides valuable inspiration for incorporating semantic relationships into our proposed interactive visualization system.

Prior research has contributed foundational techniques in multimodal interaction, scalability, navigation, iterative exploration, and semantic graph analysis. Building on these works, our system integrates and extends these ideas into a unified framework for interactive exploration, refinement, and visualization of named entities across large text corpora.

\section{Design Principles}
\label{sec:requirements}

In this section, we highlight key limitations of existing named entity visualization systems and draw from the broader visualization literature to derive five core design principles for our approach:

\begin{itemize}
  \setlength\itemsep{1em}
    \item \textbf{Integrated data exploration and visualization --} Graph-based visualizations of named entities often involve selecting a subset of entities and relations from a larger corpus, the scope and relevance of which may not be known in advance. This selection process can become labor-intensive if performed manually. Existing network visualization systems often assume that all entities and relations in a dataset are to be included in the visualization, overlooking the exploratory step required to identify subsets of interest. This can result in cluttered or unreadable graphs, as well as delays in achieving meaningful insights.
    
    \item \textbf{Cross-Document Entity Relations --}
    Beyond the classification of individual mentions, \ac{nel} and coreference extraction are often employed to link multiple entity instances that refer to the same underlying entity. This task extends both within a document, where entities may appear under different surface forms, abbreviations, or titles, and across documents, where mentions must be linked despite variation in spelling, context, or language. Commonly used named entity visualizations, such as displaCy~\cite{Honnibal2020} (see~\cref{fig:data_model} (left), struggle to showcase such relations effectively as they are only depicted as properties of individual entity instances.
    
    \item \textbf{Retaining the Fuzziness of Entity Data --} Information extraction processes related to named entities, such as \ac{ner} and \ac{nel}, rarely yield unambiguous results, especially for historical and heterogeneous corpora, where metadata is often incomplete~\cite{Ehrmann2023}. Instead, they often produce multiple candidate entities and relations with associated confidence scores~\cite{Hu2024,Zhang2024}. Instead of only preserving the ``best guess'', we aim to showcase multiple candidates and let the researcher explore the spectrum of possible interpretations.
    
    \item \textbf{Optional automation --} While automation is critical for making graph visualizations accessible, it should not eliminate authorial control. Meaningful network visualizations often require manual refinement, such as pruning peripheral nodes, re-weighting edges, or emphasizing particular subgraphs. Automation should therefore serve as an initial scaffold, enabling authors to iteratively adjust and enhance the visualization to highlight the most relevant aspects of the data.
    
    \item \textbf{Progressive visualization --} Interactive graph exploration depends heavily on timely and comprehensible visual feedback. Research indicates that user interactions should produce visual responses within 50–100 ms to maintain a fluid sense of control~\cite{Liu2014}. At the same time, sudden or large-scale layout changes can disorient users and obscure the perceived impact of their actions. Progressive visualization techniques~\cite{Angelini2018} mitigate these issues by subdividing expensive computations, such as force-directed layouts or entity disambiguation steps, into incremental updates with smooth transitions~\cite{Heer2007}. This approach provides clarity and responsiveness while supporting iterative exploration of large or evolving networks.
\end{itemize}
\section{System Design}
\label{sec:system-design}

NERVIS is an interactive web-based system for graph visualization and editing, implemented using HTML, CSS, TypeScript, and React framework\footnote{\url{https://react.dev}}. The implementation of the system is available at \url{https://github.com/UL-FRI-LGM/NERVIS}. The system is designed around a modular architecture that separates data management, visualization, and user interaction. At its core, NERVIS maintains structured data models that represent nodes, edges, and associated metadata, enabling efficient storage, retrieval, and manipulation of graph information. These models feed into a graph visualization pipeline, which computes layouts, applies styling and interaction rules, and renders the graph for exploration and analysis.

The user interface provides direct access to the system’s functionality, allowing users to inspect and modify graph elements, control visualization parameters, and apply filters. The interface is organized into distinct components that facilitate navigation, node-level editing, and rule-based filtering, providing a cohesive environment for interactive graph analysis.

\subsection{Data Model}
\label{sec:data-model}

In this section, we present the data model underlying NERVIS, designed according to the principles of \emph{Cross-Document Entity Relations} and \emph{Retaining the Fuzziness of Entity Data} (see~\cref{sec:requirements}).

\begin{figure}[h]
    \centering
    \includegraphics[width=\linewidth]{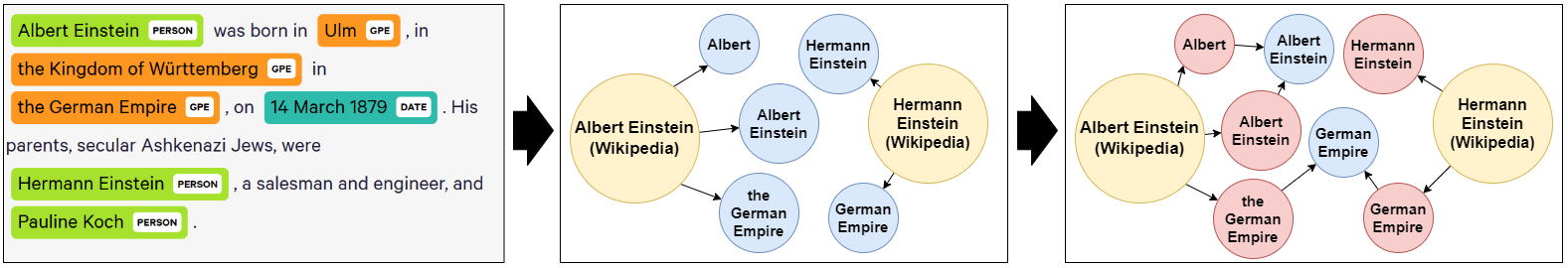}
    \caption{Data model diagrams for text-based named entity visualizations from displaCy~\cite{Honnibal2020} (left), simple document-entity model (middle), and proposed document-mention-entity model (right).}
    \label{fig:data_model}
\end{figure}

Typical text-based named entity visualizations represent entities as annotated spans over a sequence of tokens (see \cref{fig:data_model}, left). To extend this approach to a network graph spanning multiple documents, entities are first reformulated as \emph{entity instance nodes}, while token sequences are abstracted into \emph{document nodes} (\cref{fig:data_model}, middle). Although this representation captures basic entity-document relationships, it cannot effectively represent higher-level relations, such as links between mentions of the same entity across different documents. To address this, entity instance nodes are further separated into \emph{mention nodes}, representing individual occurrences of an entity within a document, and \emph{entity nodes}, representing the underlying entity itself (\cref{fig:data_model}, right).

Both \emph{entity} and \emph{mention nodes} include an entity class property derived from the CoNLL-2003 dataset~\cite{Sang2003}, which classifies named entities as \emph{Person}, \emph{Organization}, \emph{Location}, or \emph{Miscellaneous}. This classification was chosen because it underpins many state-of-the-art \ac{ner} methods~\cite{Hu2024}. To maintain the principle of data fuzziness, mentions are not strictly constrained to match the class of their connected entity, enabling the system to highlight potential inconsistencies while allowing users to correct them at their discretion.

The data model also defines four types of connections \ie graph edges. First, \emph{mention-to-document} and \emph{mention-to-entity} connections reflect relationships provided in the input data, and as per design principles, impose no limit on the number of entities a mention may be associated with. Second, \emph{mention-collocation} connections capture which mentions occur within the same sentence. Finally, a virtual \emph{entity-to-document} connection is introduced to facilitate observation of cross-document entity relations and to enable more efficient filtering and exploration of the graph.

\subsection{Visualization Pipeline}
\label{sec:visulization-pipeline}

\begin{figure}[h]
    \centering
    \includegraphics[width=1.0\linewidth]{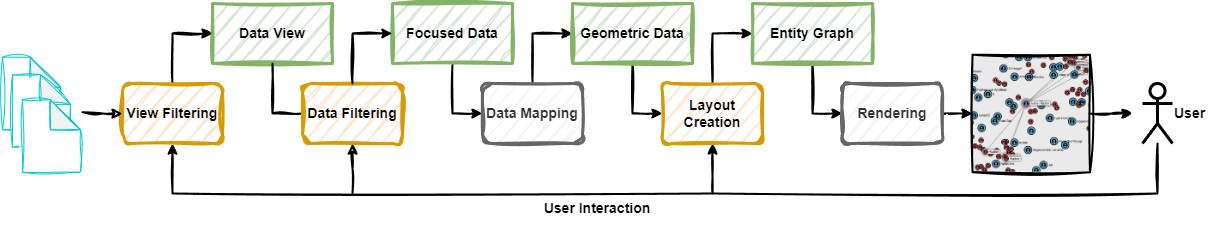}
    \caption{NERVIS visualization pipeline. Orange and gray nodes indicate interactable and uninteractable stages, respectively. Green nodes indicate intermediary data representations.}
    \label{fig:pipeline}
\end{figure}

The NERVIS visualization pipeline is illustrated in \cref{fig:pipeline} and is modeled after the classical visualization pipeline framework~\cite{Card1999}. It is designed to transform the structured data described in~\cref{sec:data-model} into an interactive graph representation, and is organized into five main stages:

\begin{enumerate}
    \item \textbf{View Filtering --} The structural abstraction of the graph is adjusted to emphasize different levels of detail in the visualization.
    \item \textbf{Data Filtering --} Input data is refined according to user-specified criteria, such as rule-based filters, user selection, or document subsets, to focus the visualization on relevant nodes and edges.
    \item \textbf{Data Mapping --} Filtered data is mapped onto geometric structures, with selected elements optionally receiving additional visual emphasis based on user interactions.
    \item \textbf{Layout Creation --} Node and edge positions are computed to produce a spatial arrangement that highlights relational structures and preserves graph readability.
    \item \textbf{Rendering} - The graph is visually rendered on the display.
\end{enumerate}

\subsubsection{View Filtering}
\label{sec:view-selection}

The first stage of the pipeline allows users to toggle between two structural views of the graph: the default Document–Mention–Entity (D–M–E) view and a simplified Document–Entity (D–E) view. In the D–E view, mention nodes are abstracted away and replaced with virtual entity-to-document edges, producing a more concise visualization that emphasizes cross-document entity relations.

This choice directly shapes the set of nodes and edges made available to subsequent stages of the pipeline. By selecting the D–M–E view, users retain the full detail of entity mentions within documents, enabling fine-grained exploration of textual annotations. By contrast, the D–E view allows users to focus on higher-level relationships across documents without the visual clutter of mention nodes.

\begin{figure}[h!]
    \fbox{\includegraphics[width=0.48\linewidth]{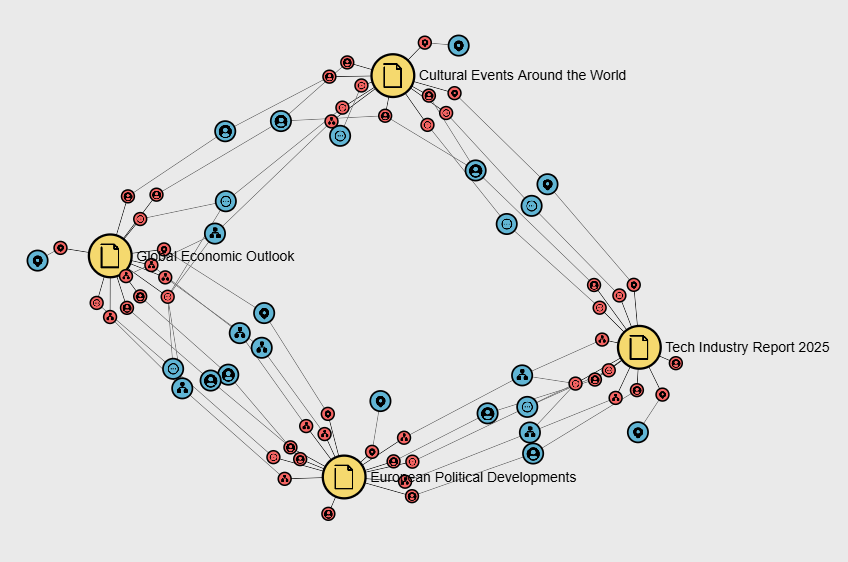}}%
    \hfill
    \fbox{\includegraphics[width=0.48\linewidth]{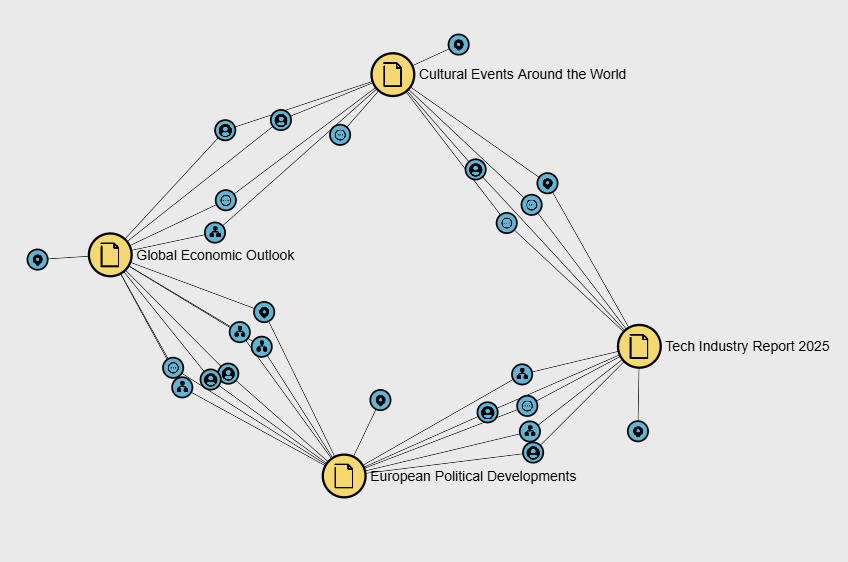}}%
    \caption{A comparison of Document–Mention–Entity view (left) and Document–Entity view (right).}
    \label{fig:two_images}
\end{figure}

\subsubsection{Data Filtering}
\label{sec:data-filtering}

\begin{figure}[h!]
    \includegraphics[width=\linewidth]{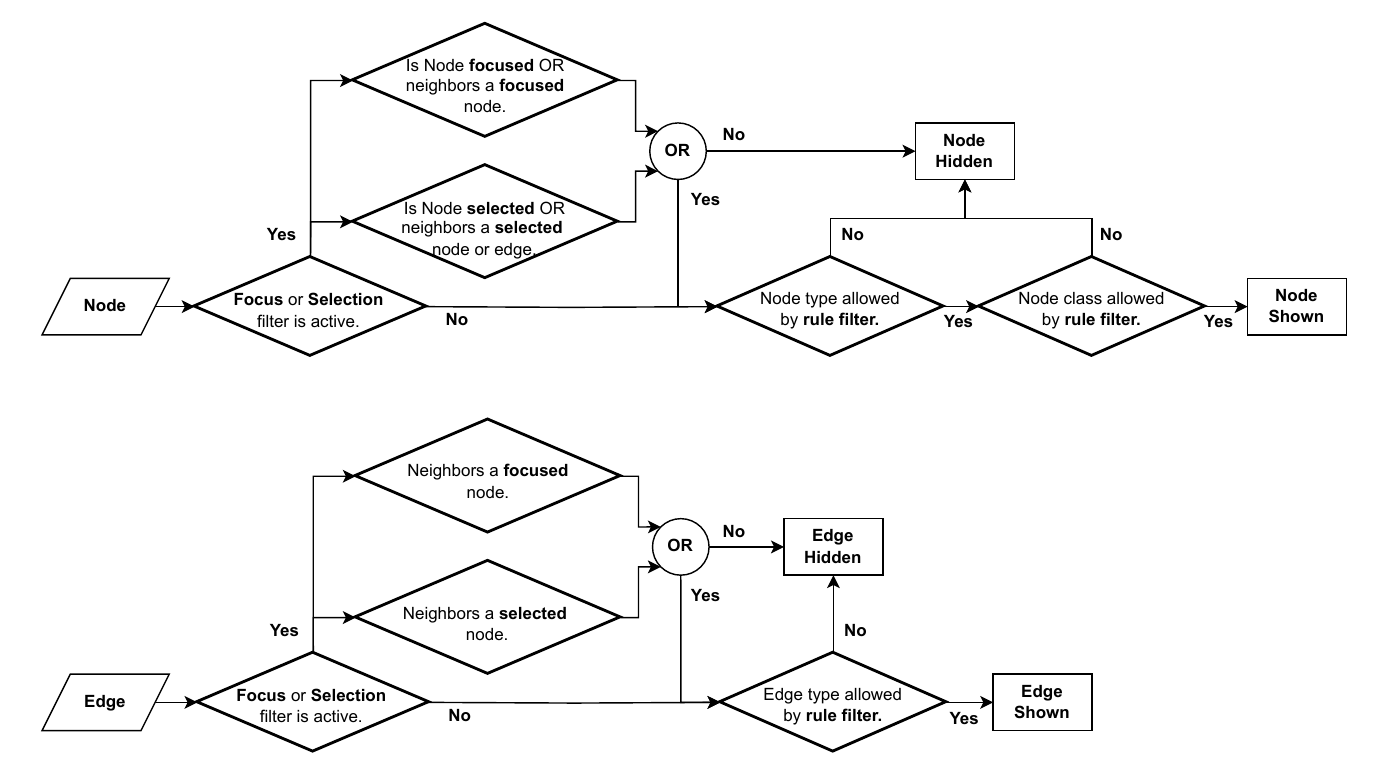}
    \caption{Node (top) and edge (bottom) filtering algorithm flowchart.}
    \label{fig:algorithms}
\end{figure}

To satisfy the \emph{Integrated Data Exploration and Visualization} design principle (see \cref{sec:requirements}), NERVIS supports interactive filtering of the input data. Inspired by other network visualization tools (\eg PatientFlow~\cite{Kwak2023} and CellNetVis~\cite{Heberle2017}), we implement two filtering steps:
\begin{enumerate}
  \setlength\itemsep{1em}
    \item \textbf{Focus filter --} This filter allows the user to specify a node or edge of interest. Two focus filters can be applied concurrently: the \emph{selection filter} and the \emph{node focus filter}. The \emph{selection filter} highlights a chosen node by removing all edges that do not connect it to its neighbors, reducing ambiguity in dense graphs. The \emph{node focus filter} removes all nodes and edges not directly connected to a specified node, enabling users to concentrate on the local neighborhood of interest. If applied concurrently, they are designed not to conflict with each other; the selection filter will not remove edges of the focused node and \textit{vice versa}.
    
    \item \textbf{Rule filter --} This filter allows users to control which types of nodes and edges are displayed based on their properties. Filtering can be performed by node type (e.g., \emph{document}, \emph{mention}, \emph{entity}), by entity class (e.g., \emph{Person}, \emph{Location}), or by any combination of these criteria.
\end{enumerate}
The complete filtering algorithm is illustrated in~\cref{fig:algorithms}.

\subsubsection{Data Mapping}
\label{sec:data-mapping}

In the data mapping stage, the filtered data is transformed into geometric structures that will be rendered in the final visualization. This process is subject to two key constraints. First, the number of nodes can reach several thousand, requiring the use of simple and recognizable shapes to maintain clarity in high-density graphs. Second, the layout algorithms, described in~\cref{sec:layout-creation}, are similarly limited to simple geometric structures to ensure computational efficiency and maintain performance.

\begin{figure}[h]
    \centering
    \includegraphics[width=0.4\linewidth]{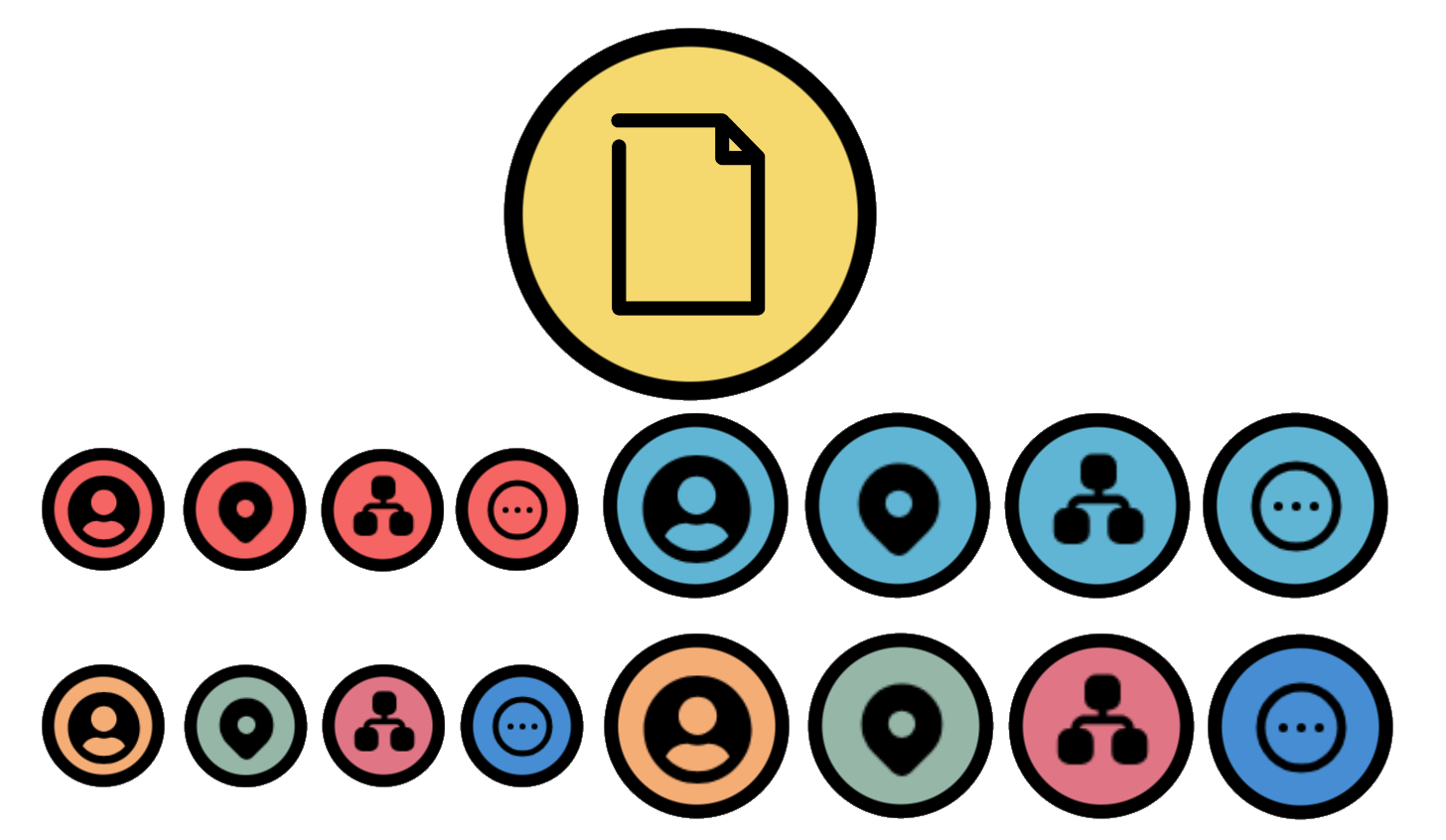}
    \caption{Visual structures used to depict different node types. The top row shows a \emph{document} node. The middle row illustrates \emph{mention} nodes (left) and \emph{entity} nodes (right) across different entity classes. The bottom row depicts mention and entity nodes using an entity class-based color scheme. From left to right the individual node pictograms depict \emph{person}, \emph{location}, \emph{organization} and \emph{miscellaneous} entity classes.}
    \label{fig:circles}
\end{figure}

To satisfy these constraints, each node is represented as a circle, with size and color encoding its type. To support a variety of use cases, two color schemes are provided: one based on node type and another based on entity class. Additionally, to enhance visual distinction and improve user interpretation, a pictogram indicating the node’s class and type is displayed. Examples of the resulting node structures are shown in~\cref{fig:circles}.

\subsubsection{Layout Creation}
\label{sec:layout-creation}

In the layout creation step, the positions of nodes within the graph are determined. In line with the design principle of \emph{optional automation} (see~\cref{sec:requirements}), NERVIS allows users to either employ an automated layout algorithm or manually adjust node positions. For automated layout generation, we use the \emph{ForceAtlas2} algorithm~\cite{Jacomy2014}, which is efficient and supports iterative computation, aligning with our design principle of \emph{progressive visualization}.

We employ the implementation provided by the Graphology framework~\cite{Guillaume2025}, which leverages the Web Workers API\footnote{\url{https://html.spec.whatwg.org/multipage/workers.html}} to multithread the computation, ensuring that the user interface remains responsive during layout processing. An example of a layout produced by the ForceAtlas2 algorithm is shown in~\cref{fig:gui}.

\subsubsection{Rendering}
\label{sec:rendering}

Rendering constitutes the final stage of the visualization pipeline, responsible for producing the visual representation of the graph on the user interface. To ensure a smooth and responsive experience when visualizing high-density graphs, NERVIS integrates the \emph{Sigma.js}\footnote{\url{https://github.com/jacomyal/sigma.js}} framework. Sigma.js employs an instance-based WebGL rendering pipeline, allowing computations to be offloaded to the GPU and fully leveraging modern graphics hardware for high-performance visualization.

One limitation of this rasterized rendering approach is that it does not produce vector-based outputs, which restricts the possibility of external post-processing or vector editing of the visualization. However, this is compensated by a comprehensive set of in-system editing tools, including node manipulation, focus and rule filtering, and layout adjustments, which allow users to modify the graph without requiring external editing.

\subsection{User Interface}
\label{sec:gui}

\begin{figure}[h]
    \centering
    \includegraphics[width=\linewidth]{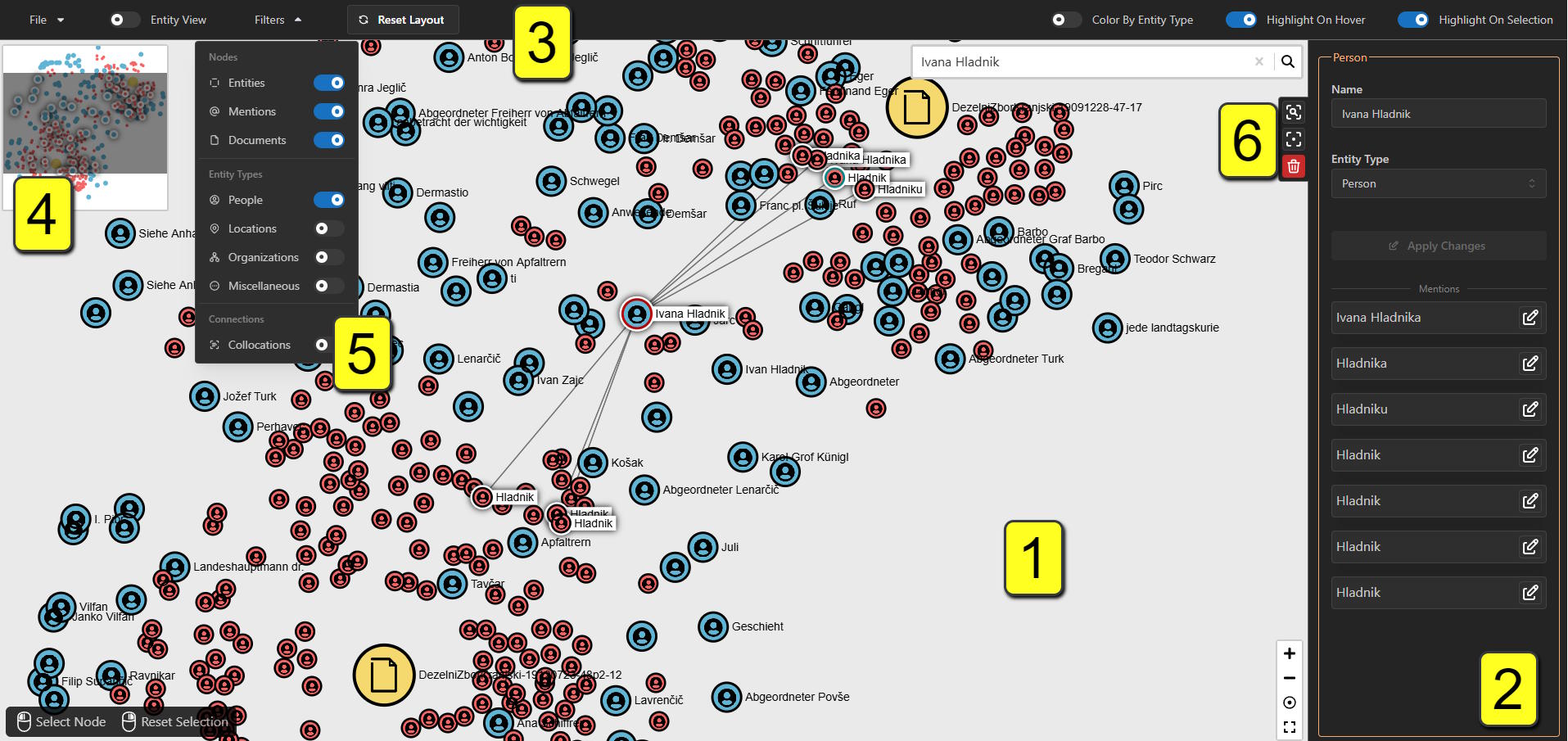}
    \caption{Overview of NERVIS graphical user interface. (1) Graph rendered using the visualization pipeline described in~\cref{sec:visulization-pipeline}. (2) Node editing widget (3) Toolbar. (4) Graph's minimap. (5) Selection of rule filters. (6) Node-specific actions and filters. }
    \label{fig:gui}
\end{figure}

The NERVIS system provides a user interface for interactive graph exploration and editing (see~\cref{fig:gui}). The main display (1) presents the graph produced by the visualization pipeline described in~\cref{sec:visulization-pipeline}, enabling detailed inspection of nodes and edges. Additionally, the event-based interaction system allows direct selection, movement, and deletion of nodes and edges on the main display. The node editing widget (2) allows direct modification of node attributes and properties within the interface, and also supports immediate update of the visualized nodes.

The toolbar (3) offers core workflow functions, including graph import and export, view toggling, and activation of focus filters to highlight or hide specific portions of the graph. The toolbar also allows for efficient search and selection of nodes within the graph, utilizing MiniSearch\footnote{\url{https://github.com/lucaong/minisearch}}, an efficient full-text search engine for JavaScript. A minimap (4) provides an overview of the graph layout, assisting in navigation and orientation on larger graphs. The rule filter panel (5) permits selective filtering according to user-defined criteria, allowing attention to be directed toward relevant patterns and relationships. The node actions panel (6) allows access to three context-sensitive operations: contextual zoom, which sets the view to encompass the node and its visible neighbors, the activation of the node focus filter (see~\cref{sec:data-filtering}), and node deletion.

\section{Expert Evaluation}
\label{sec:expert-evaluation}

Because NERVIS builds on established technologies, such as the Sigma.js rendering framework for GPU-accelerated graph drawing, a dedicated performance evaluation was not conducted. Instead, we carried out an expert evaluation to assess the system as a whole, focusing on usability, interaction design, and the extent to which the implemented features support the intended workflows.

We discussed the resulting system with a \ac{dh} scholar studying historical print media, investigating relationships and representations of historical figures and geographic locations. Since systems for \ac{ner} and \ac{nel} perform worse than humans, historians have learned to be very skeptical of purely computer-based approaches. In order to make historical research reliable, software for interacting with computer-produced results is a critical part of the \ac{dh} research process.

After initial consultation set the stage and expectations, we performed two rounds of expert system evaluation with synthetic data. After preliminary evaluation, the feature requests were prioritized, and the most important ones were implemented. After the final round, the expert judged the system to be mature enough for use in a real-life setting. They also composed the list of possible future improvements, additions, and extensions.

In the initial consultation, we discussed software goals and previous experience when handling named-entity data with existing software. One of the key things pointed out during this consultation was the burden of human labor when dealing with computer-generated labels. For historians, a recall-focused approach is preferred to an accuracy-focused one. They prefer broad sets of result candidates, which they can browse through and edit manually as opposed to only seeing a single most likely result. This encouraged the design choice to develop a tool where the deletion of nodes and edges is the most common user operation, as opposed to creating new ones. A high recall approach inevitably generates relatively high amounts of noise, which we planned to tackle with a variety of content filters that allow researchers to focus on specific documents, entities, or types when cleaning the data. Since the users of \ac{dh} tools often do not have an engineering background and often struggle with the installation process, we decided on the web platform, which also makes the system broadly available.

After the prototype was tested, we followed up with our second consultation. At this stage, the tool received generally positive feedback, with feature requests focusing on the most common user actions. As already mentioned, with high recall results, most user actions involve deleting nodes and edges, which was made additionally accessible with the implementation of hotkeys for quick node/edge deletion. We discussed the noisiness of the visualization when there are many documents present, which exposed the need for users to focus on a single document when cleaning the data. While collocations are a common tool for studying text, in this particular task, the question of which entities appear in the same sentence is of relatively limited importance. We decided to prioritize edges between documents, mentions, and entities while initially hiding the edges denoting collocations.

The final evaluation presented us with a new list of requirements and feature requests, which are the subject of future work. The need for undo/redo actions and action history was expressed alongside the wish to see specific mentions in the text context. Finally, the need to merge recognized entity nodes arose due to errors in the \ac{nel} process.
\section{Conclusion}
\label{sec:conclusion}

In this paper, we introduced an interactive web-based system for the visualization and editing of named entity graphs. The system builds on existing technologies such as Graphology~\cite{Guillaume2025} and Sigma.js, but contributes a tailored data model, a structured visualization pipeline, and interaction techniques specifically designed for large-scale named entity graphs.

Because the system leverages established rendering and layout technologies, we concentrated our evaluation on usability and interaction rather than raw performance. To this end, we conducted an expert evaluation with a digital humanities scholar working on historical print media. The evaluation provided valuable feedback on design decisions, particularly regarding the trade-off between recall and noise in named-entity data, the need for efficient editing operations, and the importance of content filters to reduce cognitive load when exploring large graphs. Iterative consultation shaped the prototype into a system the expert deemed ready for real-world use, while also identifying directions for further development, such as undo/redo functionality, contextual access to mentions, and support for merging entity nodes.

\section*{Acknowledgment}
We would like to thank Filip Dobranić for his expert feedback in evaluation of our system. This work was supported by the Slovenian Research and Innovation Agency research programme ``Digital Humanities: resources, tools and methods'' (2022--2027) [grant number P6-0436] and by the project ``Large Language Models for Digital Humanities'' (2024--2027) [grant number GC-0002].

\bibliography{literature}

\begin{thebibliography}{22}
\expandafter\ifx\csname natexlab\endcsname\relax\def\natexlab#1{#1}\fi
\providecommand{\url}[1]{\texttt{#1}}
\providecommand{\href}[2]{#2}
\providecommand{\path}[1]{#1}
\providecommand{\DOIprefix}{doi:}
\providecommand{\ArXivprefix}{arXiv:}
\providecommand{\URLprefix}{URL: }
\providecommand{\Pubmedprefix}{pmid:}
\providecommand{\doi}[1]{\href{http://dx.doi.org/#1}{\path{#1}}}
\providecommand{\Pubmed}[1]{\href{pmid:#1}{\path{#1}}}
\providecommand{\bibinfo}[2]{#2}
\ifx\xfnm\relax \def\xfnm[#1]{\unskip,\space#1}\fi
\bibitem[{Paul et~al.(2018)Paul, Chang, Endert, Cramer, Gillen, Hampton, Burtner, Perko, and Cook}]{Paul2018}
\bibinfo{author}{C.~L. Paul}, \bibinfo{author}{J.~S. Chang}, \bibinfo{author}{A.~Endert}, \bibinfo{author}{N.~Cramer}, \bibinfo{author}{D.~Gillen}, \bibinfo{author}{S.~Hampton}, \bibinfo{author}{R.~Burtner}, \bibinfo{author}{R.~Perko}, \bibinfo{author}{K.~Cook},
\newblock \bibinfo{title}{{TexTonic: Interactive Visualization for Exploration and Discovery of Very Large Text Collections}},
\newblock \bibinfo{journal}{{Information Visualization}}  (\bibinfo{year}{2018}). \DOIprefix\doi{10.1177/1473871618785390}.
\bibitem[{Lee et~al.(2012)Lee, Isenberg, Riche, and Carpendale}]{Lee2012}
\bibinfo{author}{B.~Lee}, \bibinfo{author}{P.~Isenberg}, \bibinfo{author}{N.~H. Riche}, \bibinfo{author}{S.~Carpendale},
\newblock \bibinfo{title}{{Beyond Mouse and Keyboard: Expanding Design Considerations for Information Visualization Interactions}},
\newblock \bibinfo{journal}{{IEEE Transactions on Visualization and Computer Graphics}}  (\bibinfo{year}{2012}). \DOIprefix\doi{10.1109/tvcg.2012.204}.
\bibitem[{Tao et~al.(2019)Tao, Liu, Wang, Battle, Demiralp, Chang, and Stonebraker}]{Tao2019}
\bibinfo{author}{W.~Tao}, \bibinfo{author}{X.~Liu}, \bibinfo{author}{Y.~Wang}, \bibinfo{author}{L.~Battle}, \bibinfo{author}{c.~Demiralp}, \bibinfo{author}{R.~Chang}, \bibinfo{author}{M.~Stonebraker},
\newblock \bibinfo{title}{{Kyrix: Interactive Pan/Zoom Visualizations at Scale}},
\newblock \bibinfo{journal}{{Computer Graphics Forum}}  (\bibinfo{year}{2019}). \DOIprefix\doi{10.1111/cgf.13708}.
\bibitem[{Al~Nasar et~al.(2011)Al~Nasar, Mohd, and Ali}]{AlNasar2011}
\bibinfo{author}{M.~R. Al~Nasar}, \bibinfo{author}{M.~Mohd}, \bibinfo{author}{N.~M. Ali},
\newblock \bibinfo{title}{{A Conceptual Framework for an Interactive Personal Information Management System}}  (\bibinfo{year}{2011}). \DOIprefix\doi{10.1109/iuser.2011.6150545}.
\bibitem[{Sedig and Parsons(2013)}]{Sedig2013}
\bibinfo{author}{K.~Sedig}, \bibinfo{author}{P.~Parsons},
\newblock \bibinfo{title}{{Interaction Design for Complex Cognitive Activities With Visual Representations: A Pattern-Based Approach}},
\newblock \bibinfo{journal}{{AIS Transactions on Human-Computer Interaction}}  (\bibinfo{year}{2013}). \DOIprefix\doi{10.17705/1thci.00055}.
\bibitem[{Satyanarayan et~al.(2016)Satyanarayan, Russell, Hoffswell, and Heer}]{Satyanarayan2016}
\bibinfo{author}{A.~Satyanarayan}, \bibinfo{author}{R.~P. Russell}, \bibinfo{author}{J.~Hoffswell}, \bibinfo{author}{J.~Heer},
\newblock \bibinfo{title}{{Reactive Vega: A Streaming Dataflow Architecture for Declarative Interactive Visualization}},
\newblock \bibinfo{journal}{{IEEE Transactions on Visualization and Computer Graphics}}  (\bibinfo{year}{2016}). \DOIprefix\doi{10.1109/tvcg.2015.2467091}.
\bibitem[{Rahman et~al.(2021)Rahman, Islam, Akter, Akter, Islam, and Xu}]{Rahman2021}
\bibinfo{author}{M.~F. Rahman}, \bibinfo{author}{M.~R. Islam}, \bibinfo{author}{S.~Akter}, \bibinfo{author}{S.~Akter}, \bibinfo{author}{L.~Islam}, \bibinfo{author}{G.~Xu},
\newblock \bibinfo{title}{{DiaVis: Exploration and Analysis of Diabetes Through Visual Interactive System}},
\newblock \bibinfo{journal}{{Human-Centric Intelligent Systems}}  (\bibinfo{year}{2021}). \DOIprefix\doi{10.2991/hcis.k.211025.001}.
\bibitem[{Heberle et~al.(2017)Heberle, Carazzolle, Telles, Meirelles, and Minghim}]{Heberle2017}
\bibinfo{author}{H.~Heberle}, \bibinfo{author}{M.~F. Carazzolle}, \bibinfo{author}{G.~P. Telles}, \bibinfo{author}{G.~V. Meirelles}, \bibinfo{author}{R.~Minghim},
\newblock \bibinfo{title}{Cellnetvis: a web tool for visualization of biological networks using force-directed layout constrained by cellular components},
\newblock \bibinfo{journal}{BMC Bioinformatics} \bibinfo{volume}{18} (\bibinfo{year}{2017}) \bibinfo{pages}{395}. \DOIprefix\doi{10.1186/s12859-017-1787-5}.
\bibitem[{Kwak et~al.(2023)Kwak, Park, and Song}]{Kwak2023}
\bibinfo{author}{K.~Kwak}, \bibinfo{author}{J.~Park}, \bibinfo{author}{H.~Song},
\newblock \bibinfo{title}{A visual analytics framework for inter-hospital transfer network of stroke patients},
\newblock \bibinfo{journal}{Applied Sciences} \bibinfo{volume}{13} (\bibinfo{year}{2023}). \DOIprefix\doi{10.3390/app13095241}.
\bibitem[{Rabzelj et~al.(2023)Rabzelj, Bohak, Ju\v{z}ni\v{c}, Kos, and Sedlar}]{Rabzelj2023}
\bibinfo{author}{M.~Rabzelj}, \bibinfo{author}{C.~Bohak}, \bibinfo{author}{L.~v. Ju\v{z}ni\v{c}}, \bibinfo{author}{A.~Kos}, \bibinfo{author}{U.~Sedlar},
\newblock \bibinfo{title}{{Cyberattack Graph Modeling for Visual Analytics}},
\newblock \bibinfo{journal}{{IEEE Access}} \bibinfo{volume}{11} (\bibinfo{year}{2023}) \bibinfo{pages}{86910--86944}. \DOIprefix\doi{10.1109/ACCESS.2023.3304640}.
\bibitem[{Mongiov\`{i} and Gangemi(2024)}]{Mongiovi2024}
\bibinfo{author}{M.~Mongiov\`{i}}, \bibinfo{author}{A.~Gangemi},
\newblock \bibinfo{title}{{GRAAL: Graph-Based Retrieval for Collecting Related Passages Across Multiple Documents}},
\newblock \bibinfo{journal}{{Information}}  (\bibinfo{year}{2024}). \DOIprefix\doi{10.3390/info15060318}.
\bibitem[{Honnibal et~al.(2020)Honnibal, Montani, Van~Landeghem, and Boyd}]{Honnibal2020}
\bibinfo{author}{M.~Honnibal}, \bibinfo{author}{I.~Montani}, \bibinfo{author}{S.~Van~Landeghem}, \bibinfo{author}{A.~Boyd},
\newblock \bibinfo{title}{{spaCy: Industrial-strength Natural Language Processing in Python}}  (\bibinfo{year}{2020}). \DOIprefix\doi{10.5281/zenodo.1212303}.
\bibitem[{Ehrmann et~al.(2023)Ehrmann, Hamdi, Pontes, Romanello, and Doucet}]{Ehrmann2023}
\bibinfo{author}{M.~Ehrmann}, \bibinfo{author}{A.~Hamdi}, \bibinfo{author}{E.~L. Pontes}, \bibinfo{author}{M.~Romanello}, \bibinfo{author}{A.~Doucet},
\newblock \bibinfo{title}{Named entity recognition and classification in historical documents: A survey},
\newblock \bibinfo{journal}{ACM Comput. Surv.} \bibinfo{volume}{56} (\bibinfo{year}{2023}). \DOIprefix\doi{10.1145/3604931}.
\bibitem[{Hu et~al.(2024)Hu, Hou, and Liu}]{Hu2024}
\bibinfo{author}{Z.~Hu}, \bibinfo{author}{W.~Hou}, \bibinfo{author}{X.~Liu},
\newblock \bibinfo{title}{{Deep learning for named entity recognition: a survey}},
\newblock \bibinfo{journal}{{Neural Computing and Applications}} \bibinfo{volume}{36} (\bibinfo{year}{2024}) \bibinfo{pages}{8995--9022}. \DOIprefix\doi{10.1007/s00521-024-09646-6}.
\bibitem[{Zhang et~al.(2024)Zhang, Zhao, Gao, and Hu}]{Zhang2024}
\bibinfo{author}{Z.~Zhang}, \bibinfo{author}{Y.~Zhao}, \bibinfo{author}{H.~Gao}, \bibinfo{author}{M.~Hu},
\newblock \bibinfo{title}{{LinkNER: Linking Local Named Entity Recognition Models to Large Language Models using Uncertainty}},
\newblock in: \bibinfo{booktitle}{{Proceedings of the ACM Web Conference 2024}}, WWW '24, \bibinfo{publisher}{Association for Computing Machinery}, \bibinfo{address}{New York, NY, USA}, \bibinfo{year}{2024}, p. \bibinfo{pages}{4047–4058}. \DOIprefix\doi{10.1145/3589334.3645414}.
\bibitem[{Liu and Heer(2014)}]{Liu2014}
\bibinfo{author}{Z.~Liu}, \bibinfo{author}{J.~Heer},
\newblock \bibinfo{title}{{The Effects of Interactive Latency on Exploratory Visual Analysis}},
\newblock \bibinfo{journal}{{IEEE TVCG}} \bibinfo{volume}{20} (\bibinfo{year}{2014}) \bibinfo{pages}{2122--2131}. \DOIprefix\doi{10.1109/TVCG.2014.2346452}.
\bibitem[{Angelini et~al.(2018)Angelini, Santucci, Schumann, and Schulz}]{Angelini2018}
\bibinfo{author}{M.~Angelini}, \bibinfo{author}{G.~Santucci}, \bibinfo{author}{H.~Schumann}, \bibinfo{author}{H.-J. Schulz},
\newblock \bibinfo{title}{{A Review and Characterization of Progressive Visual Analytics}},
\newblock \bibinfo{journal}{{Informatics}} \bibinfo{volume}{5} (\bibinfo{year}{2018}). \DOIprefix\doi{10.3390/informatics5030031}.
\bibitem[{Heer and Robertson(2007)}]{Heer2007}
\bibinfo{author}{J.~Heer}, \bibinfo{author}{G.~Robertson},
\newblock \bibinfo{title}{{Animated Transitions in Statistical Data Graphics}},
\newblock \bibinfo{journal}{{IEEE TVCG}} \bibinfo{volume}{13} (\bibinfo{year}{2007}) \bibinfo{pages}{1240--1247}. \DOIprefix\doi{10.1109/TVCG.2007.70539}.
\bibitem[{Tjong Kim~Sang and De~Meulder(2003)}]{Sang2003}
\bibinfo{author}{E.~F. Tjong Kim~Sang}, \bibinfo{author}{F.~De~Meulder},
\newblock \bibinfo{title}{{Introduction to the {C}o{NLL}-2003 Shared Task: Language-Independent Named Entity Recognition}},
\newblock in: \bibinfo{booktitle}{{Proceedings of the Seventh Conference on Natural Language Learning at {HLT}-{NAACL} 2003}}, \bibinfo{year}{2003}, pp. \bibinfo{pages}{142--147}. \URLprefix \url{https://www.aclweb.org/anthology/W03-0419}.
\bibitem[{Card et~al.(1999)Card, Mackinlay, and Shneiderman}]{Card1999}
\bibinfo{editor}{S.~K. Card}, \bibinfo{editor}{J.~D. Mackinlay}, \bibinfo{editor}{B.~Shneiderman} (Eds.), \bibinfo{title}{{Readings in information visualization: using vision to think}}, \bibinfo{publisher}{Morgan Kaufmann Publishers Inc.}, \bibinfo{address}{San Francisco, CA, USA}, \bibinfo{year}{1999}.
\bibitem[{Jacomy et~al.(2014)Jacomy, Venturini, Heymann, and Bastian}]{Jacomy2014}
\bibinfo{author}{M.~Jacomy}, \bibinfo{author}{T.~Venturini}, \bibinfo{author}{S.~Heymann}, \bibinfo{author}{M.~Bastian},
\newblock \bibinfo{title}{{ForceAtlas2, a Continuous Graph Layout Algorithm for Handy Network Visualization Designed for the Gephi Software}},
\newblock \bibinfo{journal}{{PLOS ONE}} \bibinfo{volume}{9} (\bibinfo{year}{2014}) \bibinfo{pages}{1--12}. \DOIprefix\doi{10.1371/journal.pone.0098679}.
\bibitem[{Plique(2025)}]{Guillaume2025}
\bibinfo{author}{G.~Plique}, \bibinfo{title}{{Graphology, a robust and multipurpose Graph object for JavaScript.}}, \bibinfo{year}{2025}. \DOIprefix\doi{10.5281/zenodo.14835805}.

\end{thebibliography}

\end{document}